\documentclass{elsart}
\usepackage{graphicx}
\usepackage{amssymb}

\begin{document}
\begin{frontmatter}
 
\title{The Use of the Monte Carlo Hamiltonian}

\author[LU]{H. Kr\"oger\corauthref{cor}},
\author[CCAST,ZSU]{X.Q. Luo},
\author[DU]{K.J.M. Moriarty}

\address[LU]{D\'epartement de Physique, Universit\'e Laval, 
Qu\'ebec, Qu\'ebec G1K 7P4, Canada}

\address[ZSU]{Department of Physics,
       Zhongshan University, Guangzhou 510275, 
       China}

\address[CCAST]{CCAST (World Laboratory), P.O. Box 8730, 
Beijing 100080, China}

\address[DU]{Department of Mathematics, 
Statistics and Computational Science, 
Dalhousie University, Halifax, Nova Scotia B3H 3J5, Canada}

\corauth[cor]{Corresponding author. E-mail: hkroger@phy.ulaval.ca}

\begin{abstract}
In order to solve quantum field theory in a non-perturbative way, Lagrangian lattice simulations have been very successful. Here we discuss a recently proposed alternative Hamiltonian lattice formulation - the Monte Carlo Hamiltonian. In order to show its working in the case of the scalar $\Phi^{4}_{1+1}$ model, we have computed thermodynamic functions like free energy, average energy, entropy and specific heat. We find good agreement between the results from the Monte Carlo Hamiltonian and standard Lagrangian lattice computations. However, the Monte Carlo Hamiltonian results 
show less fluctuations under variation of temperature. 
We address properties of the MC Hamiltonian, like a finite temperature window, and scaling properties. Also we discuss possible future applications - like 
quantum chaos in many-body systems, the non-perturbative computation of wave functions of elementary particles, as well as scattering amplitudes in high energy physics.
\end{abstract}

\end{frontmatter}

\section{Introduction}
\label{sec:Introduction}
Lagrangian lattice field theory has been for the last three decades the most successful non-perturbative technique to compute properties of elementary particles and to solve $QCD$. 
However, some problems have resisted to a solution even by this powerful technique. For example, it is difficult to estimate wave functions and 
the spectrum of excited states. 
Wave functions in conjunction with the energy spectrum 
contain more physical information than the energy spectrum alone. 
Although lattice $QCD$ simulations in the Lagrangian formulation
give good estimates of the hadron masses,
one is yet far from a comprehensive understanding of
hadrons. Let us take as example a new type of hardrons made of
gluons, the so-called glueballs. 
Lattice $QCD$ calculations\cite{kn:Luo96}
predict the mass of the lightest glueball 
with quantum number $J^{PC}=0^{++}$,
to be $1650 \pm 100 MeV$. Experimentally, there are at least two
candidates: $f_0(1500)$ and $f_J(1710)$.
The investigation of the glueball production and decays  
can certainly provide additional important information for
experimental determination of a glueball. 
Therefore, it is important to be able to compute the glueball wave function. 

On the other hand, the Hamiltonian formulation seems very suitable to compute the energy spectrum and wave function of the ground state and excited states.
Unfortunately, the Hamiltonian formulation applied to many-body systems or elementary particle physics has been widely considered to suffer from the absence of a solid many-body solution technique. There are a few exceptions: In 1-dimension, the density matrix renormalisation group technique works very well \cite{kn:White}.
In Ref.\cite{kn:Jirari99,kn:Huang02} we have made a proposal how to overcome this problem, i.e. by constructing an effective Hamiltonian via Monte Carlo - the so called Monte Carlo Hamiltonian (MCH).

In recent years, Monte Carlo methods have been widely used to solve problems in quantum physics. 
For example, with quantum Monte Carlo there has been improvement in nuclear shell model calculations \cite{kn:Otsuka}. 
A proposal to solve the sign problem in Monte Carlo Greens function method, 
useful for spin models has been made by Sorello \cite{kn:Sorello}.
Lee et al. \cite{kn:Lee} have suggested a method to diagonalize Hamiltonians, 
via a random search of basis vectors having a large overlap with low-energy eigenstates. 

In this article we will not review the mathematical formulation defining the 
Monte Carlo Hamiltonian, but rather discuss the underlying physical idea. Also, we will examine some of its characteristic properties, namely 
the fact that the MCH describes physics in a window of energy and temperature.
We ask - which are the parameters which determine the position and size of the window? How can it be widened? Also we examine a scaling property -
although the MCH is calculated at a certain inverse temperature $\beta_0$, it 
holds for all $\beta$ in the respective window.  
Finally, we discuss avenues of possible future application of the MCH -
like quantum chaos in many-body systems of quantum field theory,
wave functions in elementary particle physics, and scattering in particle physics.    

\begin{figure}[thb]
\begin{center}
\includegraphics[scale=0.6,angle=270]{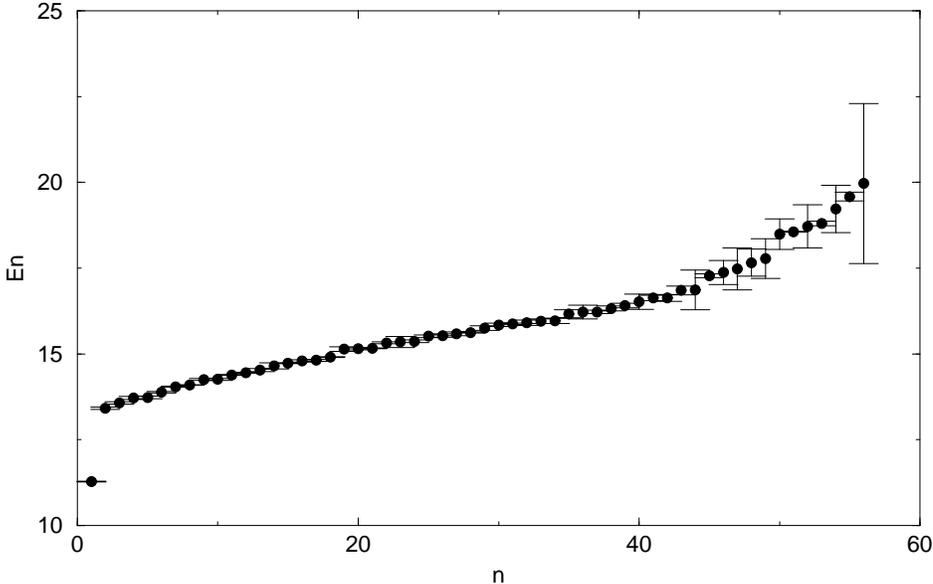}
\end{center}
\caption{Energy spectrum in a low energy window.}
\label{Fig1}
\end{figure}

\section{Monte Carlo Hamiltonian - an effective Hamiltonian}
\label{sec:Effective Hamiltonian}

The use of Hamiltonian methods in many-body physics has a long history.
In general, the Hamiltonian is an operator - expressed in terms of fields and its conjugate momenta. A widely adopted strategy to compute physics from such Hamiltonian proceeds via construction of a Fock-space, on which such Hamiltonian operates, and to construct a finite dimensional model subspace (e.g. Tamm-Dancoff approximation). By computing matrix elements from such subspace, 
one computes a finite dimensional matrix and obtains - via diagonalisation - a spectrum of energies and wave functions. The problem with such procedure is that  a cut-off in Fock space particle number and in high momentum is uncontrolled.
A much better idea is the renormalisation group idea by Kadanoff and Wilson.
They suggest to construct (via blockspin transformation or intregrating out high momentum components) a Hamiltonian for the physics of critical phenomena,
valid in the neighborhood of a fixed point - which needs much less degrees of freedom than the original Hamiltonian. 

The Monte Carlo Hamiltonian is similar to this in the sense that it aims 
to decribe physics only in a finite range of parameters (i.e. a low energy window). In lattice field theory, one constructs a Hamiltonian via the transfer matrix - i.e. going to the limit of transition time $T \to 0$.
This leads - for lattice gauge theory - to the well known Kogut-Susskind Hamiltonian, which is expressed in terms of gauge fields (link variables) and its conjugate momenta. In contrast, the MC Hamiltonian is obtained from transition matrix elements at finite transition time. As a result, the MC Hamiltonian is obtained and defined in terms of its matrix elements - and not in terms of operators. As pointed out above, the crucial question is: In which subspace or model space are those matrix elements to be taken? Contrary to the 
Tamm-Dancoff Hamiltonian expressed in terms of particle creation and annihilation operators in momentum space, the MC Hamiltonian is formulated in a space, which is the analogue of position space in quantum mechanics - the so called Bargman space. It is interesting to note that the problem to conserve manifestly local gauge invariance in lattice gauge - apparently necessary for confinement of quarks - has been solved by Wilson, using a position space formulation (e.g. Wilson loops) and not a momentum space formulation. It looks like as if the description of gauge theory in nature is simpler in position space. 

However, the use of Bargman space is by itself not good enough to solve a many-body problem. An idea - essential for the MC Hamiltonian - is to use a model subspace created in a random way (Monte Carlo), however, guided by a physical principle, which gives a large ponderation to such states which are expected to yield "large" (non-negligeable) transition matrix elements. 
This idea has been taken over from Lagrangian lattice field theory, where
the solution of path integrals - via Monte Carlo with importance sampling -
allows to compute (estimate) physical observables by summing over field configurations $\Phi$ which give a "large" weight $\exp[-S(\Phi)/\hbar]$
where $S$ is the action. In other words, we suggest that Monte Carlo with importance sampling, which allows to solve lattice field theory in the Lagrangian formulation, is also the key to solve lattice field theory in the Hamiltonian formulation. This generates what we call an (importance sampled) stochastic basis in Bargman space, which is absolutely essential for the solution of the Hamiltonian many-body problem.
 
\begin{figure}[thb]
\begin{center}
\includegraphics[scale=0.6,angle=270]{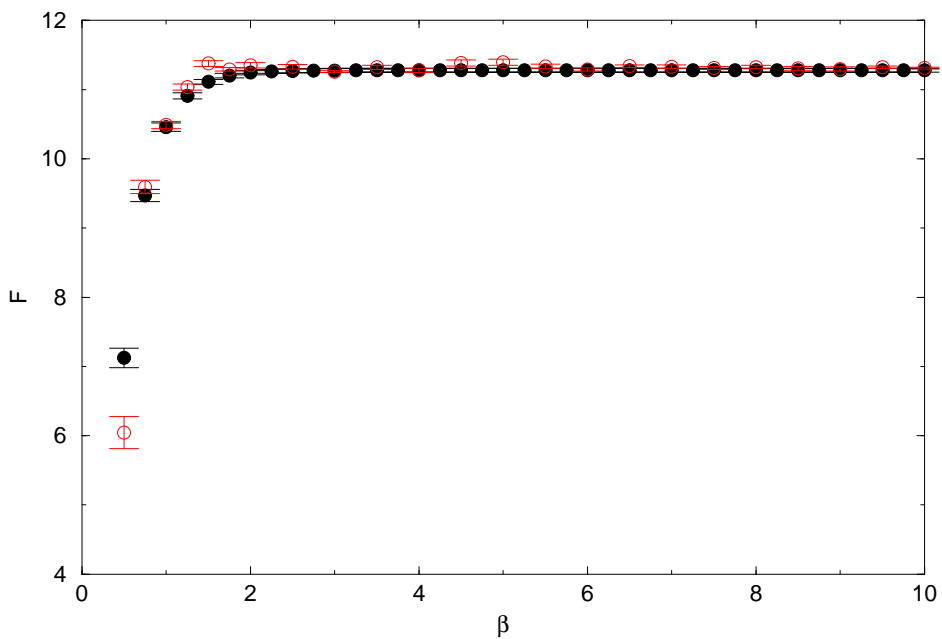}
\end{center}
\caption{Free energy $F({\beta})$. 
Comparison of results from Monte Carlo Hamiltonian (filled circles) with standard Lagrangian lattice calculations
(open circles).}
\label{Fig2}
\end{figure}

\section{Model: chain of coupled anharmonic oscillators}
\label{sec:Model}
We consider a one-dimensional chain of $N_{s}$ coupled harmonic oscillators, 
with anharmonic perturbation. Its Euclidean action is given by
\begin{eqnarray}
\label{action_osc}
S = \int dt \sum_{n=1}^{N_{s}} ~ \frac{1}{2} \dot{\phi}_{n}^{2}
+ \frac{\Omega^{2}}{2} (\phi_{n+1} - \phi_{n})^{2} 
+ \frac{\Omega_{0}^{2}}{2} \phi_{n}^{2} 
+ \frac{\lambda}{2} \phi_{n}^{4} ~ .
\end{eqnarray}
In the continuum formulation it corresponds to the scalar $\Phi^{4}_{1+1}$ model,  
\begin{eqnarray}
S = \int dt \int dx ~
\frac{1}{2} (\frac{\partial \Phi}{\partial t})^{2} 
+ \frac{1}{2} (\nabla_{x} \Phi)^{2} 
+ \frac{m^{2}}{2} \Phi^{2}
+ \frac{g}{4!} \Phi^{4} ~ .
\end{eqnarray}
Introducing a space-time lattice with lattice spacing $a_{s}$ and $a_{t}$, this action becomes
\begin{eqnarray}
\label{action_scalar}
S &=& \sum_{n=1}^{N_{s}} \sum_{k=0}^{N_{t}-1} a_{t} a_{s} 
\left[ 
\frac{1}{2} \left( \frac{ \Phi(x_{n},t_{k+1}) - \Phi(x_{n},t_{k}) }{a_{t}} \right)^{2}  \right.
\nonumber \\
&+&  \left. \frac{1}{2} \left( \frac{ \Phi(x_{n+1},t_{k}) - \Phi(x_{n},t_{k}) }{a_{s}} \right)^{2}
+ \frac{m^{2}}{2} \Phi^{2}(x_{n},t_{k})
+ \frac{g}{4!} \Phi^{4}(x_{n},t_{k}) \right] ~ .
\end{eqnarray}
The actions given by Eq.(\ref{action_scalar}) and Eq.(\ref{action_osc})
can be identified by posing $\phi=\sqrt{a_s}\Phi$, $\Omega=1/a_s$,
$\Omega_0=m$, and $\lambda/2=g/4!$.

\begin{figure}[thb]
\begin{center}
\includegraphics[scale=0.6,angle=270]{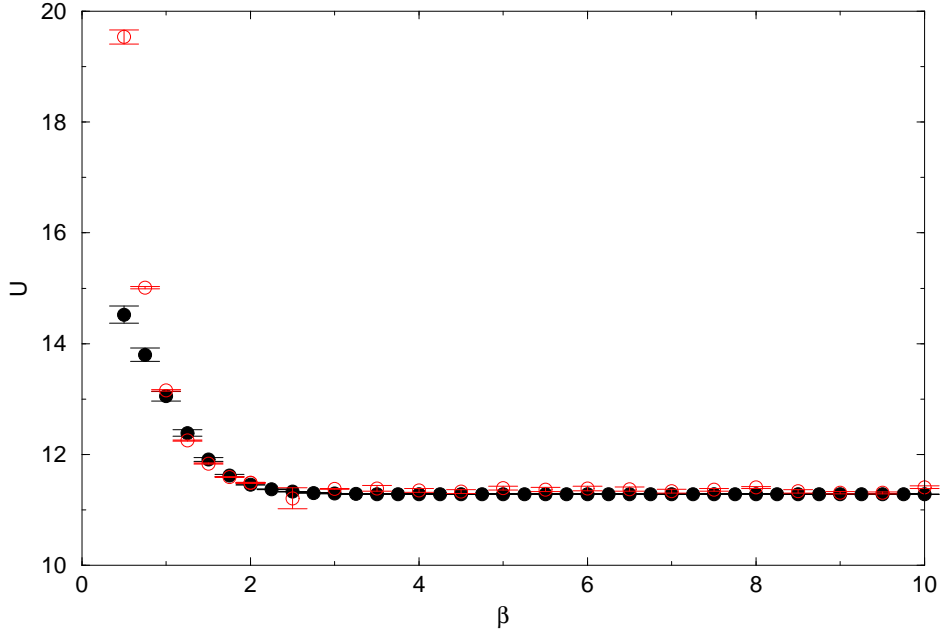}
\end{center}
\caption{Same as Fig. \ref{Fig2}, for average energy $U({\beta})$.}
\label{Fig3}
\end{figure}

\section{Comparison: Monte Carlo Hamiltonian vs. Lagrangian lattice simulation}
\label{sec:Comparison}
{\it Energy spectrum}. In Ref.\cite{kn:KleinGordon} we have tested the method for a chain of harmonic oscillators
(Klein Gordon model) where the energy spectrum is known exactly.
We found quite good agreement between the results from MC Hamilton and the exact results. The energy spectrum for the chain of anharmonic oscillators (scalar model) obtained from the MC Hamiltonian is shown in Fig.[\ref{Fig1}].
The model parameters are: $\Omega=1$, $\Omega_0=2$, $\lambda=1$ ($\hbar=1$, $k_B=1$). Approximation parameters are: Lattice size $N_s=9$, lattice spacing $a_s=1$, dimension of stochastic basis $N_{stoch}=100$. 
The statistical errors (indicated by error bars) are reasonably small and are more or less of the same size for $n=1$ to $n=40$. They increase noticeably for $n > 40$. Also the shape of the curve $E_n$ vs. $n$ changes beyond $n=40$. This signals the occurence of a low-energy window stretching from $E_{1} = 11.27$ to $E_{40} = 16.52$.

\begin{figure}[thb]
\begin{center}
\includegraphics[scale=0.6,angle=270]{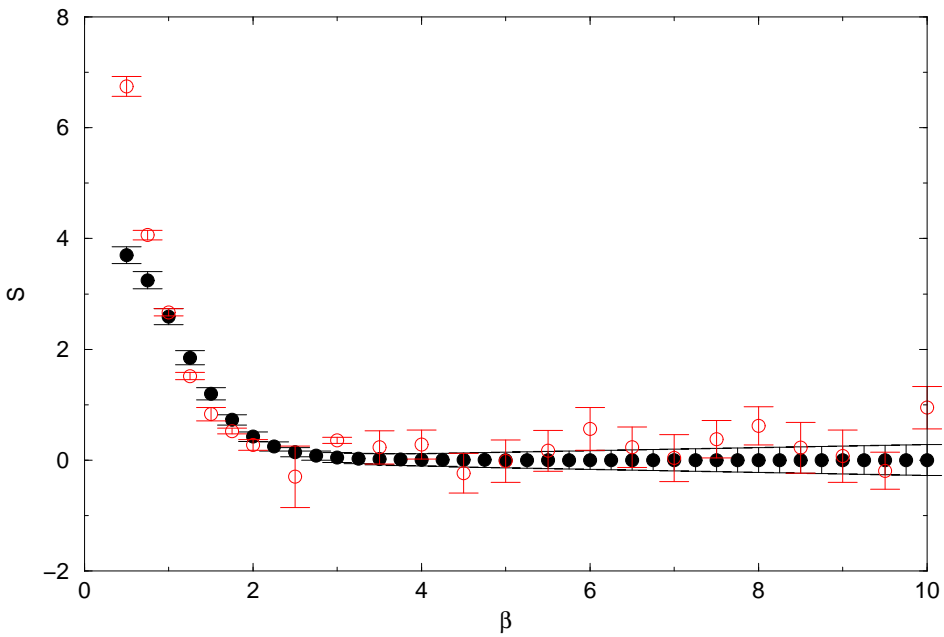}
\end{center}
\caption{Same as Fig. \ref{Fig2}, for entropy $S({\beta})$.}
\label{Fig4}
\end{figure}

{\it Thermodynamical observables}. 
A solid test of the Monte Carlo Hamiltonian method is a comparison 
with results from standard Lagrangian lattice calculations. 
The information of the energy spectrum enters into such thermodynamical functions. Thus thermodynamical functions allow an indirect test of the energy spectrum. 
We have chosen to compute the following thermodynamical observables:
the partition function $Z$, free energy $F$,
average energy $U$, specific heat $C$, entropy $S$ and pressure $P$.
Those are defined by
\begin{eqnarray}
\label{eq:ThermoDynFunc}
Z(\beta) &=& {\rm Tr} \left[ \exp \left( -\beta H \right) \right] ~ ,
\nonumber \\
F(\beta) &=& - \frac{1}{\beta} \log Z ~ ,
\nonumber \\
U(\beta) &=& {1 \over Z} {\rm Tr} \left[ H \exp \left( -\beta H \right) \right]
= - {\partial \log Z \over \partial \beta} ~ ,
\nonumber \\
C(\beta) &=& {\partial U \over \partial \tau}|_{V}
= -k_B \beta^2 {\partial U \over \partial \beta}|_{V} ~ ,
\nonumber \\
S(\beta) &=& \frac{1}{\tau}(U - F) = k_{B} \beta (U - F) ~ ,
\nonumber \\
P(\beta) &=& - {\partial F \over \partial V } ~ .
\end{eqnarray}
Here $k_B$ denotes the Boltzmann constant.  
The temperature $\tau$ is related to $\beta$ via 
$\beta =T/\hbar= 1/({k_B} \tau)$. 

\begin{figure}[thb]
\begin{center}
\includegraphics[scale=0.6,angle=270]{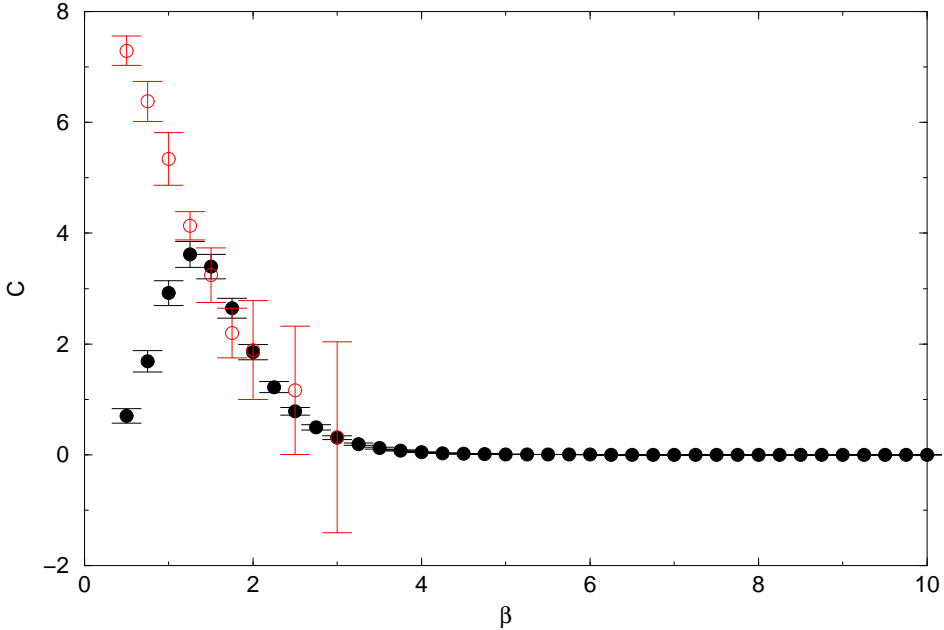}
\end{center}
\caption{Same as Fig. \ref{Fig2},  for specific heat $C({\beta})$.}
\label{Fig5}
\end{figure}

Fig.[\ref{Fig2}] shows the free energy $F$ as a function of $\beta$.
The following behavior is observed. First, there is good overall agreement in the range $0.5 \le \beta \le 10$. However, there is an essential difference between the Lagrangian approach and the MC Hamiltonian approach: 
In the Lagrangian approach, one needs to do a new computation for each value of 
temperature or inverse temperature $\beta$. Moreover, the computation is easier 
in the regime of small $\beta$, i.e. high temperature. Increasing $\beta$ requires larger lattices, which means that the numerical effort and errors increase. In the MC Hamiltonian approach, one compute transition matrix elements for a particular fixed value $\beta_{0}$. One extracts a spectrum of energy eigenvalues and corresponding wave functions. Suppose we are close to the quantum continuum limit. Then the obtained energy spectrum should become independent of $\beta_{0}$. Then plugging those energies in the expressions of thermodynamical functions, Eq.(\ref{eq:ThermoDynFunc}), allows to obtain those function for any value of $\beta$ (asymptotic scaling). Because in practice one can never reach the continuum limit nor the infinite volume limit, as a consequence one will not be able from a single MC Hamiltonian siumulation to obtain thermodynamical functions for any $\beta$. Rather one can compute those functions only in a bounded domain in $\beta$. In other words there is a finite temperature window. This is one essential difference between both approaches.
There is another difference. While the Lagrangian approach is efficient and precise in the domain of small $\beta$ or high temperature, the MC Hamiltonian approach is efficient and precise in the domain of high $\beta$ or low temperature. This can be seen in Fig.[\ref{Fig3}], which displays the average energy. With increase of $\beta$, one observes that the Lagrangian lattice results fluctuate more than those from  the MC Hamiltonian. This phenomenon is more evident in Fig.[\ref{Fig4}] for the entropy and even more so in Fig.[\ref{Fig5}] for the specific heat.  
From the data of $F$, $U$, $S$ and $C$, we estimate the temperature window 
of the Monte Carlo Hamiltonian to range from $\beta=1$ to $\beta=10$.

One may ask: On what does the location and size of the finite temperature window depend? It certainly depends on the parameters which control the approach to the continuum limit and the infinite volume limit, i.e. lattice spacing $a_{s}$ and $a_{t}$, on the lattice size $N_{s}$ and $N_{t}$. In lattice $QCD$ it will depend also on the coupling $g_{s}$ and $g_{t}$. It will depend on $N_{stoch}$, the size of the stochastic basis. Finally, it will depend also on the precision in computing the transition amplitudes (statistical error, etc.). 
From the study of the scalar model, we found that the size of the lattice in temperature direction as well the size of the stochastic basis are most important.

\section{Outlook: Possible applications of MC Hamiltonian method}
\label{sec:Outlook}
In order to make the MC Hamiltonian a valuable tool in condensed matter or elementary particle physics, it is crucial to show that it allows to treat gauge theories and eventually fermions.
Work on incorporating gauge theories is presently under way.
If the MC Hamiltonian methods works for such theories, there is much interesting physics to be done, where conventional non-perturbative methods, in particular Lagrangian lattice field theory, have brought little progress.  
Let us outline a few examples.
(i) Energies and wave functions of excited states.
Here, in our opinion, the Hamiltonian formulation is most suitable.
Many-body wave functions play a role for hadron structure functions in particle physics, electromagnetic form factors in nuclear physics, Bose-Einstein condensation in atomic physics.
(ii) S-matrix and scattering amplitudes in high energy physics.
The non-perturbative calculation of scattering amplitudes in high energy physics is an unsolved problem yet. In Ref.\cite{kn:Kroger92} Kr\"{o}ger has suggested how to compute the S-matrix in a time-dependent manner from spectral information, i.e. energies and wave functions. The MC Hamiltonian may provide such spectral information.
(iii) Quark-gluon phase transition in $QCD$ at finite temperature and finite density. Such phase transition requires to treat $QCD$ in the presence of a non-zero chemical potential. In Lagrangian Lattice QCD, the treatment of the gauge group $SU(3)$ leads to a complex action, which escapes a treatment by Monte Carlo methods \cite{kn:Montvay}. This has been a long standing problem. On the other hand, in the Hamiltonian formulation on the lattice, the presence of the chemical potential does not lead to any complex term. 
(iv) Quantum chaos from level density distributions in quantum field theories.
It is a widely used strategy to characterize quantum chaos, e.g. in nuclear or atomic physics, via level density distributions (Wigner for fully chaotic) and Poissonian (for integrabel non-chaotic) systems \cite{kn:Haake}. To obtain good statistics requires the knowledge of a considerable number of energy levels. 
The idea is that such information may be provided by the MC Hamiltonian.

\section{Summary}
\label{sec:Summary}
We have discussed physical ideas underlying the concept of the Monte Carlo Hamiltonian. For the case of the scalar model we have computed thermodynamical functions and by comparison with standard Lagrangian lattice calculations we showed that the MC Hamiltonian works. 
We outlined possible future applications of the MC Hamiltonian in elementary particle physics, nuclear physics and atomic physics.

\ack

H.K. and K.J.M.M. are grateful for support by NSERC Canada.
X.Q.L. is supported by the
National Science Fund for Distinguished Young Scholars,
National Science Foundation of China, 
the Ministry of Education, 
the Foundation of
the Zhongshan University Advanced Research Center
and the Guangdong Provincial Natural Science Foundation of China (proj. 990212).

\end{document}